# 基於 Ascon 確定性隨機位元產生器應用於嵌入式系統

Abel C. H. Chen, *Senior Member, IEEE*

*摘要*—由於確定性隨機位元產生器(Deterministic Random Bit Generator, DRBG)是隨機數產生器和密碼學應用的核心基礎，並且在資源受限的嵌入式系統(Embedded Systems)將會有記憶體空間限制和計算效能考量，所以建立高效能的安全確定性隨機位元產生器對嵌入式系統是重要的議題。除此之外，美國國家標準暨技術研究院(National Institute of Standards and Technology, NIST)於 2025 年 8 月制定 Ascon 輕量級密碼學標準，將可適用於資源受限的嵌入式系統。有鑑於此，本研究修改確定性隨機位元產生器標準，提出 Ascon 驅動的基於雜湊確定性隨機位元產生器(Ascon-Driven Hash-Based DRBG)、Ascon 驅動的金鑰雜湊訊息鑑別碼隨機位元產生器(Ascon-Driven Keyed-Hash Message Authentication Code DRBG)、以及 Ascon 驅動的計數器模式隨機位元產生器(Ascon-Driven Counter-mode DRBG)。在實驗環境，本研究實作上述方法於 Raspberry Pi，由結果顯示本研究所提方法可以比現有的確定性隨機位元產生器具有更高的計算效率，並且使用更少的記憶體空間，可適用於嵌入式系統。

*關鍵字*—確定性隨機位元產生器、雜湊、金鑰雜湊訊息鑑別碼、計數器模式、Ascon 密碼學

## I.前言

美國國家標準暨技術研究院(National Institute of Standards and Technology, NIST)於 2025 年 8 月制定 Ascon 輕量級密碼學標準[1]，並且陸續有研究開始嘗試把 Ascon 演算法應用於嵌入式系統[2]、物聯網設備和晶片設計[3]等。有鑑於此，本研究主要聚焦在把 NIST SP 800-232 定義的 Ascon 演算法[1]結合到 NIST SP 800-90A Rev. 1 定義的各種確定性隨機位元產生器(Deterministic Random Bit Generator)，包含基於雜湊確定性隨機位元產生器(Hash-Based DRBG)、金鑰雜湊訊息鑑別碼隨機位元產生器(Keyed-Hash Message Authentication Code DRBG, HMAC DRBG)、以及計數器模式隨機位元產生器(Counter-mode DRBG, CTR DRBG)[4]。

本研究主要貢獻條列如下：

- Ascon 驅動的基於雜湊確定性隨機位元產生器(Ascon-Driven Hash-Based DRBG)：首先將修改 NIST SP 800-90A Rev. 1 定義的基於湊雜函數(Hash Function, Hash_df)衍生函數(Derivation Function)[4]，把衍生函數裡的雜湊函數修改為 Ascon-Hash256，設計 Ascon 驅動的基於湊雜函數(Ascon_Hash_df)衍生函數。然後，再運用 Ascon 驅動的基於湊雜函數衍生函數作為基於雜湊確定性隨機位元產生器的核心函數。

- Ascon 驅動的金鑰雜湊訊息鑑別碼隨機位元產生器(Ascon-Driven HMAC DRBG)：首先將修改 NIST SP 800-224 ipd 定義的金鑰雜湊訊息鑑別碼[5]，把演算法裡的雜湊函數修改為 Ascon-Hash256，設計 Ascon 驅動的金鑰雜湊訊息鑑別碼(Ascon_HMAC)。然後，再運用 Ascon 驅動的金鑰雜湊訊息鑑別碼作為金鑰雜湊訊息鑑別碼隨機位元產生器的核心函數。

- Ascon 驅動的計數器模式隨機位元產生器(Ascon-Driven CTR DRBG)：首先將修改 NIST SP 800-90A Rev. 1 定義的區塊加密(Block_Encrypt)[4]，把演算法裡的進階加密標準(Advanced Encryption Standard, AES)電子密碼本模式(Electronic codebook mode, ECB mode)[6] 修改為 Ascon-AEAD128，設計 Ascon 驅動的區塊加密(Ascon_Block_Encrypt)。然後，再運用 Ascon 驅動的區塊加密作為計數器模式隨機位元產生器的核心函數。

本研究主要分為 6 節。第 II 節~第 IV 節分別介紹本研究提出的 Ascon 驅動的基於雜湊確定性隨機位元產生器(Ascon-Driven Hash-Based DRBG)、Ascon 驅動的金鑰雜湊訊息鑑別碼隨機位元產生器(Ascon-Driven HMAC DRBG)、以及 Ascon 驅動的計數器模式隨機位元產生器(Ascon-Driven CTR DRBG)。第 V 節說明實驗環境，並且驗證計算效能。最後第 VI 節總結本研究貢獻，以及討論未來研究方向。

## II. ASCON 驅動的基於雜湊確定性隨機位元產生器

本節分別在第 II.A 節介紹 Ascon 驅動的基於湊雜函數(Ascon_Hash_df)衍生函數、第 II.B 節介紹 Ascon 驅動的基於雜湊確定性隨機位元產生器。

### *A. 基於湊雜函數衍生函數*

本節修改 NIST SP 800-90A Rev. 1 定義的基於湊雜函數(Hash Function, Hash_df)衍生函數(Derivation Function)[4]，把原本的 Hash 函數修改為 NIST SP 800-232 定義的 Ascon-Hash256 函數[1]，並且可以確定 Ascon-Hash256 函數的輸出為 256 個位元，再把修改後函數命名

Abel C. H. Chen is with the Information & Communications Security Laboratory, Chunghwa Telecom Laboratories, Taoyuan 326, Taiwan (e-mail: chchen.scholar@gmail.com).

為 **Ascon_Hash_df**，如 **Algorithm 1** 所示。其中，輸入為 *inputString* 和 *outputLength*，分別作為待雜湊字串(to-be-hashed string)和輸出位元數長度。以及輸出為 *status* 和 *requestedBits*，分別表示計算狀態和結果。根據雜湊計算結果的位元字串，取左邊 *outputLength* 個位元值作為輸出結果。

---
**Algorithm 1 Ascon_Hash_df**(*inputString*, *outputLength*)
---
**Input**:
    *inputString*: The string to be hashed.
    *outputLength*: The number of bits to be returned.

**Output**:
    *status*: The status shows **SUCCESS** or **ERROR_FLAG**.
    *requestBits*: The results of **Ascon_Hash_df**.

**Ascon_Hash_df Process**:
1: *temp* = the Null string
2: *len* = $\lceil \frac{outputLength}{256} \rceil$
3: *counter* = 0x01
4: **for** *i* = 1 to *len* **do**
4.1:  *temp* = *temp* || **Ascon-Hash256**(*counter*||*outputLength*||*inputString*)
4.2:  *counter* = *counter* + 1
5: **end**
6: *requestedBits* = **leftmost**(*temp*, *outputLength*)
7: **return** (**SUCCESS**, *requestedBits*)

---

*B. 基於雜湊確定性隨機位元產生器*

為建立基於雜湊確定性隨機位元產生器，可以先根據 NIST SP 800-90A Rev. 1 定義的 Hash_DRBG_Instantiate_algorithm 產生初始 *workingState*，包含 *V*、*C*、以及 *reseedCounter*。當 *reseedCounter* 超過 $2^{48}$ 時，可以根據 NIST SP 800-90A Rev. 1 定義的 Hash_DRBG_Reseed_algorithm 產生更新 *workingState* [4]。

本節修改 NIST SP 800-90A Rev. 1 定義的 Hashgen 函數[4]，把原本的 Hash 函數修改為 NIST SP 800-232 定義的 Ascon-Hash256 函數[1]，再把修改後函數命名為 **Ascon_Hashgen**，如 **Algorithm 2** 所示。其中，輸入為 *outputLength* 和 *V*，以及輸出為 *requestedBits*。根據雜湊計算結果的位元字串，取左邊 *outputLength* 個位元值作為輸出結果。

---
**Algorithm 2 Ascon_Hashgen**(*outputLength*, *V*)
---
**Input**:
    *outputLength*: The number of bits to be returned.
    *V*: The current value of *V*.

**Output**:
    *requestBits*: The results of **Ascon_Hashgen**.

**Ascon_Hashgen Process**:
1: *m* = $\lceil \frac{outputLength}{256} \rceil$
2: *data* = *V*
3: *W* = the Null string
4: **for** *i* = 1 to *m* **do**
4.1:  *w* = **Ascon-Hash256**(*data*)
4.2:  *W* = *W* || *w*
4.3:  *data* = (*data* + 1) mod $2^{440}$
5: **end**
6: *requestedBits* = **leftmost**(*W*, *outputLength*)
7: **return** (*requestedBits*)

---

之後修改 NIST SP 800-90A Rev. 1 定義的 Hash_DRBG_Generate_algorithm [4]，把原本的 Hash 函數修改為 NIST SP 800-232 定義的 Ascon-Hash256 函數[1]，以及把 Hashgen 函數修改為 **Ascon_Hashgen** 函數(如 **Algorithm 2** 所示)，再把修改後函數命名為 **Ascon_Hash_DRBG_Generate_algorithm**，如 **Algorithm 3** 所示。其中，輸入為 *workingState*、*outputLength*、*addInput*，分別為當下的 *workingState*、輸出位元數長度、以及額外資訊。輸出為 *status*、*requestedBits*、*newWorkingState*，分別表示計算狀態、輸出結果、以及更新後的 *workingState*。

---
**Algorithm 3 Ascon_Hash_DRBG_Generate_algorithm**
                    (*workingState*, *outputLength*, *addInput*)
---
**Input**:
    *workingState*: The current values for *V*, *C*, and *reseedCounter*.
    *outputLength*: The number of bits to be returned.
    *addInput*: The additional input string received from the application.

**Output**:
    *status*: The status shows **SUCCESS** or **ERROR_FLAG**.
    *requestBits*: The results of **Ascon_Hash_DRBG_Generate_algorithm**.
    *newWorkingState*: The new values for *V*, *C*, and *reseed_reseedCounter*.

**Ascon_Hash_DRBG_Generate_algorithm Process**:
1: **if** *reseedCounter* > $2^{48}$, then **return** an indication that a reseed is required.
2: **if** *addInput* ≠ *Null*, then **do**
2.1:  *w* = **Ascon-Hash256**(0x02||*V*||*addInput*)
2.2:  *V* = (*V* + *w*) mod $2^{440}$
3: *requestedBits* = **Ascon_Hashgen**(*outputLength*, *V*)
4: *H* = **Ascon-Hash256**(0x03||*V*)
5: *V* = (*V* + *H* + *C* + *reseedCounter*) mod $2^{440}$
6: *reseedCounter* = *reseedCounter* + 1
7: *newWorkingState* = (*V*, *C*, *reseedCounter*)
8: **return** (**SUCCESS**, *requestedBits*, *newWorkingState*)

---

III. Ascon 驅動的金鑰雜湊訊息鑑別碼隨機位元產生器

本節分別在第 III.A 節介紹 Ascon 驅動的金鑰雜湊訊息鑑別碼、第 III.B 節介紹 Ascon 驅動的金鑰雜湊訊息鑑別碼隨機位元產生器。

*A. 金鑰雜湊訊息鑑別碼*

本節修改 NIST SP 800-224 ipd 定義的金鑰雜湊訊息鑑別碼[5]，把原本的 Hash 函數修改為 NIST SP 800-232 定義的 Ascon-Hash256 函數[1]，再把修改後演算法命名為 Ascon 驅動的金鑰雜湊訊息鑑別碼(Ascon_HMAC)，如 **Figure 1** 所示。其中，輸入為金鑰 *k* 和訊息 *M*，以及輸出為 256-bit 長度的雜湊值。其中，採用 NIST SP 800-224 ipd 定義的金鑰處理流程根據金鑰 *k* 產製 $K_0$，並且採用 NIST SP 800-224 ipd 定義的 ipad 和 opad 參數值[5]。

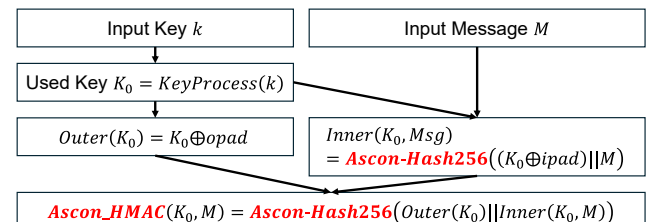

**Fig. 1.** Ascon 驅動的金鑰雜湊訊息鑑別碼

## B. 金鑰雜湊訊息鑑別碼隨機位元產生器

為建立金鑰雜湊訊息鑑別碼隨機位元產生器，可以先根據 NIST SP 800-90A Rev. 1 定義的 HMAC_DRBG_Instantiate_algorithm 產生初始 *workingState*，包含 *V*、*K*、以及 *reseedCounter*。當 *reseedCounter* 超過 $2^{48}$ 時，可以根據 NIST SP 800-90A Rev. 1 定義的 HMAC_DRBG_Reseed_algorithm 產生更新 *workingState* [4]。

本節修改 NIST SP 800-90A Rev. 1 定義的 HMAC_DRBG_Update 函數[4]，把原本的金鑰雜湊訊息鑑別碼修改為 Ascon 驅動的金鑰雜湊訊息鑑別碼(如 **Figure 1** 所示)，再把修改後函數命名為 **Ascon_HMAC_DRBG_Update**，如 **Algorithm 4** 所示。其中，輸入為 *providedData*、*K*、*V*，以及輸出為更新後的 *K* 和 *V*。

**Algorithm 4 Ascon_HMAC_DRBG_Update**(*providedData*, *K*, *V*)

**Input**:
  *providedData*: The data to be used.
  *K*: The current value of *K*.
  *V*: The current value of *V*.

**Output**:
  *K*: The new value of *K*.
  *V*: The new value of *V*.

**Ascon_HMAC_DRBG_Update Process**:
 1: *K* = **Ascon_HMAC**(*K*, *V*||0x00||*providedData*)
 2: *V* = **Ascon_HMAC**(*K*, *V*)
 3: **if** *providedData* = *Null*, then **return** (*K*, *V*)
 4: *K* = **Ascon_HMAC**(*K*, *V*||0x01||*providedData*)
 5: *V* = **Ascon_HMAC**(*K*, *V*)
 6: **return** (*K*, *V*)

之後修改 NIST SP 800-90A Rev. 1 定義的 HMAC_DRBG_Generate_algorithm [4]，把原本的 Hash 函數修改為 NIST SP 800-232 定義的 Ascon-Hash256 函數[1]，以及把 HMAC_DRBG_Update 函數修改為 **Ascon_HMAC_DRBG_Update** 函數(如 **Algorithm 4** 所示)，再把修改後函數命名為 **Ascon_HMAC_DRBG_Generate_algorithm**，如 **Algorithm 5** 所示。其中，輸入為 *workingState*、*outputLength*、*addInput*，分別為當下的 *workingState*、輸出位元數長度、以及額外資訊。輸出為 *status*、*requestedBits*、*newWorkingState*，分別表示計算狀態、輸出結果、以及更新後的 *workingState*。

## IV. Ascon 驅動的計數器模式隨機位元產生器

由於計數器模式隨機位元產生器的核心函數是區塊加密(Block_Encrypt)，所以第 IV.A 節將先介紹 Ascon 驅動的區塊加密(Ascon_Block_Encrypt)，然後第 IV.B 節介紹 Ascon 驅動的計數器模式隨機位元產生器。並且為簡化說明，本研究僅說明未使用衍生函數的計數器模式隨機位元產生器，但相似的方法也可以套用到使用衍生函數的計數器模式隨機位元產生器。

**Algorithm 5 Ascon_HMAC_DRBG_Generate_algorithm**
(*workingState*, *outputLength*, *addInput*)

**Input**:
  *workingState*: The current values for *V*, *K*, and *reseedCounter*.
  *outputLength*: The number of bits to be returned.
  *addInput*: The additional input string received from the application.

**Output**:
  *status*: The status shows **SUCCESS** or **ERROR_FLAG**.
  *requestBits*: The results of **Ascon_Hash_DRBG_Generate_algorithm**.
  *newWorkingState*: The new values for *V*, *K*, and reseed_reseedCounter.

**Ascon_HMAC_DRBG_Generate_algorithm Process**:
 1: **if** *reseedCounter* > $2^{48}$, then **return** an indication that a reseed is required.
 2: **if** *addInput* ≠ *Null*, then **do**
      (*K*, *V*) = **Ascon_HMAC_DRBG_Update**(*addInput*, *K*, *V*)
 3: *temp* = *Null*
 4: **while len**(*temp*) < *outputLength* **do**
 4.1:   *V* = **Ascon_HMAC**(*K*, *V*)
 4.2:   *temp* = *temp* || *V*
 5: *requestedBits* = **leftmost**(*temp*, *outputLength*)
 6: (*K*, *V*) = **Ascon_HMAC_DRBG_Update**(*addInput*, *K*, *V*)
 7: *reseedCounter* = *reseedCounter* + 1
 8: **return** (**SUCCESS**, *requestedBits*, *newWorkingState*)

### A. 區塊加密

由於 NIST SP 800-90A Rev. 1 定義的區塊加密(Block_Encrypt) 函數是採用進階加密標準(Advanced Encryption Standard, AES) 電子密碼本模式(Electronic codebook mode, ECB mode)[4],[6]，所以輸入僅有金鑰 *K* 和待加密資料 *V*。，並且輸出僅有密文，不具備標籤(tag)。然而，NIST SP 800-232 定義的 Ascon-AEAD128.enc(*K*, *N*, *A*, *V*)函數[1]則另外還帶有隨機數(nonce) *N* 和關聯資料(associated data) *A*；除此之外，Ascon-AEAD128.enc(*K*, *N*, *A*, *V*)函數的輸出帶有密文和標籤。有鑑於此，當欲把區塊加密函數改為 Ascon-AEAD128.enc(*K*, *N*, *A*, *V*)函數時，需新增隨機數 *N* 和關聯資料 *A*，並且在取得輸出時可以只取密文的部分，限定待加密資料 *V* 長度和密文長度一致，忽略標籤。

### B. 計數器模式隨機位元產生器

為建立計數器模式隨機位元產生器，可以先根據 NIST SP 800-90A Rev. 1 定義的 CTR_DRBG_Instantiate_algorithm 產生初始 *workingState*，包含 *V*、*K*、以及 *reseedCounter*。但需要特別說明的是，Ascon 驅動的計數器模式隨機位元產生器還需要隨機數 *N* 和關聯資料 *A*，所以初始 *workingState* 在此情境有 128 位元長度的 *V*、128 位元長度的 *K*、128 位元長度的 *N*、*A*、以及 *reseedCounter*。其中，隨機數 *N* 為隨機產製，而關聯資料 *A* 則是可以設定特定值。當 *reseedCounter* 超過 $2^{48}$ 時，可以根據 NIST SP 800-90A Rev. 1 定義的 CTR_DRBG_Reseed_algorithm 產生更新 *workingState* [4]。

本節修改 NIST SP 800-90A Rev. 1 定義的 CTR_DRBG_Update 函數[4]，把原本的 Block_Encrypt 函數修改為 NIST SP 800-232 定義的 Ascon-AEAD128.enc(*K*, *N*, *A*, *V*) 函數[1]，再把修改後函數命名為

基於 Ascon 確定性隨機位元產生器應用於嵌入式系統　　　　　　　　　　　　　　　　　　　　　　　　　　　　　4

**Ascon_CTR_DRBG_Update**，如 **Algorithm 6** 所示。其中，輸入為 *providedData*、*K*、*V*、*N*、*A*，以及輸出為更新後的 *K* 和 *V*。

---

**Algorithm 6 Ascon_CTR_DRBG_Update**(*providedData*, *K*, *V*, *N*, *A*)

**Input**:
　*providedData*: The 256-bit data to be used.
　*K*: The current value of *K*.
　*V*: The current value of *V*.
　*N*: The value of nonce.
　*A*: The value of associated data.

**Output**:
　*K*: The new value of *K*.
　*V*: The new value of *V*.

**Ascon_CTR_DRBG_Update Process**:
1: *temp* = *Null*
2: **while len**(*temp*) < 256 **do**
2.1:　**if** *ctrLen* < 128, then **do**
2.1.1:　　inc = (**rightmost**(*V*, *ctrLen*) + 1) mod $2^{ctrLen}$
2.1.2:　　*V* = **leftmost** (*V*, 128 – *ctrLen*) ∥ *inc*
　　**else** *V* = (*V*+1) mod $2^{128}$
2.2:　*outputBlock* = **Ascon-AEAD128.enc(*K*, *N*, *A*, *V*)**
2.3:　*temp* = *temp*∥*outputBlock*
3: *temp* = **leftmost**(*temp*, 256)
4: *temp* = *temp* ⊕ *providedData*
5: *K* = **leftmost**(*temp*, 128)
6: *V* = **rightmost**(*temp*, 128)
7: **return** (*K*, *V*)

---

之後修改 NIST SP 800-90A Rev. 1 定義的 CTR_DRBG_Generate_algorithm [4]，把原本的 Block_Encrypt 函數修改為 NIST SP 800-232 定義的 Ascon-AEAD128.enc(*K*, *N*, *A*, *V*) 函數 [1]，以及把 CTR_DRBG_Update 函數修改為 **Ascon_CTR_DRBG_Update** 函數(如 **Algorithm 6** 所示)，再把修改後函數命名為 **Ascon_CTR_DRBG_Generate_algorithm**，如 **Algorithm 7** 所示。其中，輸入為 *workingState*、*outputLength*、*addInput*，分別為當下的 *workingState*、輸出位元數長度、以及額外資訊。輸出為 *status*、*requestedBits*、*newWorkingState*，分別表示計算狀態、輸出結果、以及更新後的 *workingState*。

---

**Algorithm 7 Ascon_CTR_DRBG_Generate_algorithm**
　　　　　　　　(*workingState*, *outputLength*, *addInput*)

**Input**:
　*workingState*: The current values for *V*, *K*, *N*, *A*, and *reseedCounter*.
　*outputLength*: The number of bits to be returned.
　*addInput*: The additional input string received from the application.

**Output**:
　*status*: The status shows **SUCCESS** or **ERROR_FLAG**.
　*requestBits*: The results of **Ascon_Hash_DRBG_Generate_algorithm**.
　*newWorkingState*: The new values for *V*, *K*, and reseed_*reseedCounter*.

**Ascon_CTR_DRBG_Generate_algorithm Process**:
1: **if** *reseedCounter* > $2^{48}$, then **return** an indication that a reseed is required.
2: **if** *addInput* ≠ *Null*, then **do**
2.1:　*temp* = **len**(*addInput*)
2.2:　**if** (temp < seedlen), then **do** *addInput* = *addInput* ∥ $0^{256 - temp}$
2.3:　(*K*, *V*) = **Ascon_CTR_DRBG_Update**(*addInput*, *K*, *V*, *N*, *A*)
　　**else** *addInput* = $0^{256}$
3: *temp* = *Null*
4: **while len**(*temp*) < *outputLength* **do**
4.1:　**if** *ctrLen* < 128, then **do**
4.1.1:　　inc = (**rightmost**(*V*, *ctrLen*) + 1) mod $2^{ctrLen}$
4.1.2:　　*V* = **leftmost** (*V*, 128 – *ctrLen*) ∥ *inc*
　　**else** *V* = (*V*+1) mod $2^{128}$
4.2:　*outputBlock* = **Ascon-AEAD128.enc(*K*, *N*, *A*, *V*)**
4.3:　*temp* = *temp*∥*outputBlock*
5: *temp* = **leftmost**(*temp*, *outputLength*)
6: (*K*, *V*) = **Ascon_CTR_DRBG_Update**(*addInput*, *K*, *V*, *N*, *A*)
7: *reseedCounter* = *reseedCounter* + 1
8: **return** (**SUCCESS**, *requestedBits*, *newWorkingState*)

---

TABLE I
確定性隨機位元產生器比較

| DRBG | Cryptography | Utilized Memory (Byte) | Computation Time (ms) |
|---|---|---|---|
| Hash-Based DRBG | SHA-256 [7] | 5216 | 0.100 |
| Ascon-Driven Hash-Based DRBG | Ascon-Hash256 (The Proposed) | 2608 | 0.103 |
| HMAC DRGB | SHA-256 [8] | 10424 | 0.133 |
| Ascon-Driven HMAC DRBG | Ascon-Hash256 (The Proposed) | 5208 | 0.154 |
| CTR DRBG | AES-128 [9] | 8072 | 0.114 |
| Ascon-Driven CTR DRBG | Ascon-AEAD128 (The Proposed) | 5208 | 0.109 |

## V. 實驗結果與討論

為驗證本研究提出的基於 Ascon 確定性隨機位元產生器，本研究採用 Raspberry Pi 4 進行實驗，並且在 BouncyCastle 函式庫基礎上修改，以 Java 語言實作。其中，在分析已使用記憶體(used memory)空間是採用 java.lang.Runtime 類別衡量執行 DRBG 前後的已使用記憶體空間差異所取得。各別執行 1 萬次，每次產製 256 個隨機位元，平均計算時間如表 I 所示。由實驗結果可以觀察到，本研究所提方法可以用較少的記憶體空間，並且 Ascon 驅動的計數器模式隨機位元產生器具有比較短的計算時間。然而，由於 Ascon-Hash256 的計算時間比 SHA-256 的計算時間長，所以基於雜湊確定性隨機位元產生器有較高的計算效率；並且，由於金鑰雜湊訊息鑑別碼隨機位元產生器需要較多次的雜湊計算，所以 Ascon 驅動的金鑰雜湊訊息鑑別碼隨機位元產生器需要較多的計算時間。

## VI. 結論與未來研究

本研究提出多種基於 Ascon 確定性隨機位元產生器，並且可以用較少的記憶體空間，並且 Ascon 驅動的計數器模式隨機位元產生器具有高效率，更適用於嵌入式系統。未來可以考慮把 Ascon 確定性隨機位元產生器實作到晶片裡，並且運用硬體加速，以落地到物聯網和相關應用。

## 參考文獻